\documentclass[doublecol]{epl2} 
\usepackage{color}
\usepackage{amsmath}
\usepackage{amsfonts}
\usepackage{amsbsy}
\usepackage{mathrsfs}
\usepackage{graphicx}  
\usepackage{subfigure}
\newcommand{\eeq}{\end{equation}}
\newcommand{\bea}{\begin{eqnarray}}
	\newcommand{\eea}{\end{eqnarray}}

\def\lsim{\mathrel{\rlap{
			\lower4pt\hbox{\hskip-3pt$\sim$}}
		\raise1pt\hbox{$<$}}}     %less than approx. symbol
\def\gsim{\mathrel{\rlap{
			\lower4pt\hbox{\hskip-3pt$\sim$}}
		\raise1pt\hbox{$>$}}}     %greater than or approx. symbol

\title{Monte Carlo simulations of the electron short-range quantum ordering
in Coulomb systems: Reducing the ``fermionic sign problem''}
\shorttitle{Monte Carlo simulations of the electron short-range quantum ordering}
\author{V. Filinov\inst{1} \and P. Levashov\inst{1,2} \and A. Larkin\inst{1}}
\shortauthor{V. Filinov \etal}

\institute{                    
  \inst{1} Joint Institute for High Temperatures, Russian Academy of Sciences,
  Izhorskaya 13 bldg 2, Moscow 125412, Russia\\
  \inst{2} Moscow Institute of Physics and Technology, 9 Institutskiy per., Dolgoprudny, Moscow Region, 141700, Russia
}

\abstract{
To account for the interference effects of the Coulomb and exchange interactions of electrons  
a new path integral representation of the density matrix has been developed in the canonical ensemble at finite temperatures. 
The developed representation allows to reduce the notorious ``fermionic sign problem'' in the Monte Carlo simulations of fermionic 
systems. The obtained results for pair distribution functions in plasma and uniform electron gas demonstrate the 
short-range quantum ordering of electrons associated with exchange excitons in the literature. 
This fine physical effect was not observed earlier in standard path integral Monte Carlo simulations. 
}

\begin{document}

\maketitle

\section{Introduction}
One of the cornerstone challenges in the path integral Monte Carlo (PIMC) simulations 
of fermionic systems is the ``sign problem''. The  ``sign problem'' arises from the antisymmetrization 
of the fermion density matrix \cite{feynmanquantum,zamalin1977monte,EbelForFil,ForFilLarEbl} resulting in 
small differences of large numbers associated 
with even and odd permutations and the exponential growth of PIMC statistical errors.  
To overcome this issue a lot of approaches have been developed, 
but the ``fermionic sign problem'' for strongly correlated fermions has not been completely solved during 
the last fifty years. 
In \cite{larkin2017pauli,larkin2017peculiarities} the Wigner formulation of quantum mechanics has been used 
avoid the antisymmetrization of matrix elements and hence the ``sign problem'' and 
to realize the Pauli blocking of fermions. However this approach is not applicable at high degeneracy.   

In \cite{brown2013path} to reduce the ``fermionic sign problem'', the restricted fixed--node path--integral 
Monte Carlo (RPIMC ) simulations of the uniform electron gas (UEG) at 
finite temperature has been developed. In RPIMC to avoid the ``sign problem'' only positive permutations 
are taken into account, so the accuracy of the results is unknown and the error is difficult to quantify.  
More interesting approaches are the new permutation blocking path integral Monte Carlo (PB-PIMC) 
and the configuration path integral Monte Carlo (CPIMC) methods \cite{dornheim2018uniform}. 
The main idea of CPIMC is to evaluate the density matrix in the space of occupation numbers. 
However it turns out that the CPIMC method also exhibits the ``sign problem''.
In PB-PIMC the sum over permutations is presented in the form of a determinant, 
which can be calculated by the direct highly accurate methods of linear algebra allowing to reduce the sign problem. 
However the disadvantage of this method is that the determinants are the sign-altering functions worsening the 
accuracy of PIMC simulations and giving rise to the ``sign problem'' of determinants. 

Here to overcome the ``sign problem'' of determinants we have developed an approximation for a highly degenerate fermion system, in which  
the exchange determinant takes the form of the Gram determinant, which is always non-negative.  
To improve the accuracy of this approximation and to account for the interference effects of the Coulomb and 
exchange interactions of electrons  we have developed a new path integral representation of density matrix, 
in which interaction is included in the exchange determinant. 
Using this representation we have developed a modified fermionic path integral Monte Carlo approach  (MFPIMC)  
and have investigated a short--range quantum ordering in Coulomb systems in the canonical ensemble at finite temperatures. 
The observed ordering is caused by the interaction of electrons with positively charged exchange holes and the excluded volume effect 
resulting in the formation of exchange excitons \cite{Weisskopf:PR:1939,lowdin1955quantum,lowd}. 
The effect is confirmed by electron--electron pair distribution functions 
for two--component plasma (TCP) and uniform electron gas (UEG). Such fine effect has not been observed earlier in standard PIMC simulations. 

\textbf{Many--particle density matrix}.
Let us start from a neutral Coulomb system of quantum electrons and positive classical (for simplicity) 
charges with the  Hamiltonian, ${\hat H}={\hat K}+{\hat U}^c$, containing the kinetic ${\hat K}$ and 
Coulomb interaction energies, ${\hat U}^c = {\hat U}_{pp}^c + {\hat  U}^c_{ee} + {U}^c_{ep}$.  
Thermodynamic properties in the canoncial ensemble with a given temperature 
$T$ and fixed volume $V$ are fully described by the density operator ${\hat \rho} = e^{-\beta {\hat H}}$, with the partition function 
$    %\begin{equation}\label{q-def}
Z(N_e,N_p,V;\beta) = \frac{1}{N_e!N_p!} \sum_{\sigma}\int_V
\rm dx \,\rho(x, \sigma ;\beta),
$     %\end{equation}
where $\beta=1/k_BT$, and $\rho(x, \sigma ;\beta)=\langle x|e^{-\epsilon {\hat H}}|x \rangle$ denotes the diagonal elements of the density matrix 
in the coordinate representation. 
% at a given value $\sigma$ of the total electron spin. 
Here $x=\{x_e,x_p\}$ 
(in units of thermal wavelengths $\tilde{\lambda}_a=\sqrt{\frac{2\pi\hbar^2\beta}{m_a M}}$) are the spatial coordinates 
of electrons and positive charges  i.e. $x_a=\{x_{1,a},\ldots, x_{N_a,a}\}$ and 
$\sigma=\{\sigma_{1,e},\ldots,\sigma_{N_e,e}\}$   are the spin degrees of freedom of the electrons. 

Generally the exact matrix elements of the density matrix of interacting quantum systems is not known, but can be constructed using a path integral 
representation~\cite{feynmanquantum,zamalin1977monte,EbelForFil}  based on the operator identity,  
$ %\begin{equation}
e^{-\beta {\hat H}}= e^{-\epsilon {\hat H}}\cdot
e^{-\epsilon {\hat H}}\dots  e^{-\epsilon {\hat H}} \quad (\epsilon = \beta/M ), 
$ %\end{equation}  
that involves $M$ identical high-temperature factors with a temperature $MT$:  
%\begin{multline}
\begin{multline} 
\rho(x, \sigma ; x, \sigma ; \beta) = \langle x|e^{-\beta {\hat H}}|x \rangle \approx {} \\
\approx {}\sum_{\sigma}\sum_{P} (- 1)^{\kappa_{P}} {\cal S}(\sigma, {P}\sigma) 
	\prod_{m=0}^{M} 
	e^{-\epsilon { U_{m}}} \langle x |e^{-\epsilon {\hat K}} |{P} x  \rangle \\
	%	= \sum_{P} (-1)^{\kappa_{P}} 
	= {} \int{  dx^{(1)} \dots dx^{(M-1)}} \exp\Biggl\{-\sum_{m=0}^{M-1}\biggl[
	\pi \left|x^{(m)}-x^{(m+1)}\right|^2  \\
	+\epsilon U(x^{(m)})
	\biggr]\Biggr\}{\rm det}\|\Psi(x)\|,	
	\label{rho_s}
\end{multline} 	
%\end{multline}    
where  index $m=0, \dots, M$ labels the off--diagonal matrix elements of the high--temperature density matrices.  
Here each high-temperature factor has been presented in the form 
$\langle x^{(m)}|e^{-\epsilon {\hat H}}|x^{(m+1)}\rangle \approx
\langle x^{(m)}|e^{-\epsilon {\hat U_m}}|x^{(m+1)}\rangle \rho^{(m)}_0$, 
($  \rho^{(m)}_0=\langle x^{(m)}|e^{-\epsilon {\hat K}}|x^{(m+1)}\rangle = e^{-\pi |x^{(m)} - x^{(m+1}|^2 )}$) 
with the error of order $1/M^2$, arising from neglecting the commutator $\epsilon^2/2 \left[K,U^c\right]$. 
In the limit $M\rightarrow \infty$ the error of the whole product is equal to zero $(\propto 1/M)$,
and we have an exact path integral representation of the partition function. 
Now each particle is presented by a 
trajectory consisting of a set of $M$ coordinates $x^{(m)}$. 
The sum is over all permutations with parity $\kappa_{P}$. 
The spin gives rise to the spin part of the density matrix (${\cal S}$) with exchange effects accounted for by the permutation
operator  $P$ acting on the electron coordinates in $x^{(M)}$ and spin projections $\sigma$. 
We assume that in the thermodynamic limit the main contribution in the sum over spin variables comes from the term related
to equal number ($N_e/2$) of electrons with the same spin projection \cite{EbelForFil,ForFilLarEbl}, so  
$    {\rm det}\|\Psi(x)\| = {\rm det} \bigl\|e^{-{\pi} \left|x_{k,e}^{(M)}-x_{t,e}^{(0)}\right|^2}\bigr\|_1^{N_e/2} 
{\rm det}\bigl\|e^{-{\pi} \left|x_{k,e}^{(M)}-x_{t,e}^{(0)}\right|^2}  \bigr\|_{N_e/2}^{N_e}$. 

The density matrix elements  involve an effective pair interaction $U=\frac{1}{2}\sum_{k,t} \Phi_{kt}$, 
which is approximated by the Kelbg potential given by the expression:
\begin{eqnarray} 
%\begin{multline}
\Phi_{ab}(x_{ab};\epsilon) =
\frac{e_a e_b}{\tilde{\lambda}_{ab} x_{ab}} 
%\\ {} \times 
\left[1-e^{-x_{ab}^2} +
\sqrt{\pi} x_{ab} \left\{1-{\rm erf}(x_{ab})\right\} \right]
\nonumber
%\label{kelbg-d}
\end{eqnarray}  
%\end{multline}
Here $\tilde{\lambda}_{ab}x_{ab}=|q_{k,a}-q_{t,b}|\tilde{\lambda}_e$, $\tilde{\lambda}^2_{ab}=2\pi\hbar^2\epsilon/m_{ab}$,
$1/m_{ab}=1/m_a+1/m_b$, $\tilde{\lambda}^2_{e}=2\pi\hbar^2\epsilon/m_e$ and ${\rm erf}(x)$ is the standard error function.
At $x_{ab}\sim \lambda_e$ the Kelbg potential coincides with the Coulomb one, but 
is finite at $x_{ab}\to 0$ due to it's quantum nature \cite{zamalin1977monte,EbelForFil,ForFilLarEbl}.  

A disadvantage of Eq.~(\ref{rho_s}) is the sign--altering  determinant ${\rm det}||\Psi(x)||$, which is the reason of the  
``sign problem'' worsening the accuracy of PIMC simulations. To reduce the ``sign problem'' 
let us replace the variables of integration 
$x^{(m)}$ by $q^{(m)}$ for any given permutation $P$  by the substitution: 
$	x^{(m)} = (Px - x)\frac{m}{M}+x + q^{(m)}$ \cite{filinov2020uniform}. 
After some transformations, the diagonal matrix elements 
of the density matrix can be rewritten  in the form of path integral over \emph{``closed''}  trajectories 
$ \{q^{(0)}, \dots, q^{(M)} \}$ with $q^{(0)} =q^{(M)} =0 $  (see Appendix for details):    
%\begin{widetext}
\begin{multline} 	
%\begin{multline}
\rho(x, \sigma ; x, \sigma ; \beta)   \,
\equiv \int  dq^{(1)} \dots dq^{(M-1)} 
%\sum_{\sigma}\sum_{P} (\pm 1)^{\kappa_{P}}{\cal S}(\sigma, {P} \sigma^\prime) 
%\nonumber\\
\exp\Bigl[-U_E  \Bigr] \\ 
%%&&	\exp\Biggl[-\sum\limits_{m = 0}^{M-1} \epsilon U\big(x + q^{(m)}\big) \Biggr] \nonumber\\ 
{}\times \Biggl\{
\sum_{\sigma}\sum_{P} (\pm 1)^{\kappa_{P}} {\cal S}(\sigma, {P} \sigma)  \\ 
\times\exp\Bigl[	-\pi \frac{|P x-x|^2}{M}  - \sum\limits_{m = 0}^{M-1} \pi |\eta^{(m)}|^2 
-\Delta U_P \Bigr] \Biggr\}  	
%		\nonumber\\
%	\epsilon U\bigl((P x -x)\frac{m}{M}+x + q^{(m)}	\bigr) - 
%	\epsilon U\big(x + q^{(m)}\big) 
%nonumber\\
%\Biggr] 
\\
\approx \int   dq^{(1)} \dots dq^{(M-1)}
\exp\Bigl\{
-\sum\limits_{m = 0}^{M-1}  \pi | \eta^{(m)}|^2  
%\epsilon U\bigl(x + q^{(m)} \bigr) \bigr] 	
- U_E \Bigr\}\\
\times{\rm det}\|\psi(x)\|, 
\label{pathint_wignerfunctionint3}
%\nonumber&&	
%		\nonumber\\
%\exp\big[-\sum\limits_{m = 0}^{M-1} \epsilon U\big(x + q^{(m)}\big) \big]
\end{multline} 	%
%	\end{multline}
%\end{widetext}
where $\det\bigl\|\psi(x)\bigr\| = \det \|\phi_{kt} \bigl\|_1^{N_e/2}\times |\mathrm{det}\bigr\|\phi_{kt} \|_{N_e/2}^{N_e} $, 
\begin{multline}
%\begin{eqnarray} 	
\phi_{kt} =  \exp \{-{\pi} \left|r_{kt}\right|^2/M \} \times {} \\
\times \exp \Biggl\{-\frac{1}{2}\sum\limits_{m = 0}^{M-1}\biggl( \epsilon \Phi \Bigl(\Bigl|r_{tk}\frac{2m}{M} 
+r_{kt} + q_{kt}^{(m)}\Bigr|  \Bigr) \\
- {} \epsilon \Phi \Bigl(\Bigl|r_{kt} + q_{kt}^{(m)}\Bigr| \Bigr) \biggr)\Biggr\} 
\label{eq:phikt}
%\end{eqnarray} 	
\end{multline} 
and
\begin{eqnarray} 	
&&{U_E}=\exp\Biggl[-\sum\limits_{m = 0}^{M-1} \epsilon U\big(x + q^{(m)}\big) \Biggr] \nonumber\\ 
&&\Delta U_P = \sum\limits_{m = 0}^{M-1} \epsilon  
U\bigl((P x -x)\frac{m}{M}+x + q^{(m)}	\bigr)  -U_E.    
\nonumber
\end{eqnarray} 
Here we introduced the following notation for the  coordinate vectors:  
$%\begin{eqnarray}
\eta^{(m)} \equiv  q^{(m)} - q^{(m+1)}$, 
%	\nonumber\\ 
%q_{kt}^{(m)}= q_k^{(m)} -q_t^{(m)}=\sum_{k'=0}^{m-1}(\eta_k^{(k')} - \eta_t^{(k')}), 
%	\nonumber\\
$r_{kt}\equiv (x_{k}-x_{t})$, $(k,t=1,\dots,N_e)$. 

The approximation given by Eq.~(\ref{pathint_wignerfunctionint3}) reproduces both limits  of the 
highly degenerate and non-degenerate systems of fermions. 
In the classical limit due to the factor  $\exp[-\pi |Px-x|^2/M]$  the main contribution comes from the identical permutation and 
the differences of potential energies in the exponent of Eq.~(\ref{pathint_wignerfunctionint3})  are equal to zero. 
At the same time for highly degenerate plasma, where the thermal wavelength is larger than the average interparticle distance 
and trajectories are highly entangled the potential energy in equation like (\ref{pathint_wignerfunctionint3}) 
weakly depends on permutations, what enables to replace all permutations $P$  by the identical one $E$ 
\cite{larkin2017pauli,larkin2017peculiarities,EbelForFil,ForFilLarEbl}.     
At intermediate degeneracies the determinant in Eq.~(\ref{pathint_wignerfunctionint3}) 
accounts for the interference effects of the Coulomb and exchange interactions of electrons. It  
allows also to reduce the ``fermionic sign problem'' by making use of direct methods to calculate the determinants.  
For better understanding of the mathematical structure 
of Eq.~(\ref{pathint_wignerfunctionint3}) let us consider the example of two spinless fermions $N_e=2$  
($q_{12}^{(m)}=q_{1}^{(m)}-q_{2}^{(m)}$, $\mathrm{det}\bigl\|\psi(x)\bigr\| =1-\phi_{12}\phi_{21}$):    
\begin{multline}	
\rho(x_{1},x_{2}) 
\equiv \int  dq^{(1)} \dots dq^{(M-1)} \\
\exp\biggl[	- \sum\limits_{m = 0}^{M-1} \pi |\eta^{(m)}|^2\biggr]
\Biggl\{\exp\bigl[ - \epsilon \Phi \bigl(|r_{12}+q_{12}^{(m)}| \bigr) \Bigr] 
%\nonumber\\	
\\	
-\exp \left\{ -\frac{\pi}{M} 2 r_{12}^2 \right\}  \Phi\Bigl[ r_{12}\left(1 - 2m/M \right) + q_{12}^{(m)} \Bigr]   \biggr\}.
\label{pathint2}
\end{multline}
One can easily check that  Eq.~(\ref{pathint2}) obtained from the approximate expression 
(\ref{pathint_wignerfunctionint3}) coincides with the exact density matrix of two fermions.  
	
\textbf{Simulation details.} 
In TCP all charges are correlated due to interaction, while in our model of UEG the positive particles 
have to be uncorrelated to simulate the neutralizing background  \cite{filinov2015fermionic,filinov2020uniform}. 
Here the density of electrons is characterized by the Brueckner parameter $r_s=a/a_B$, 
where 
$a=[3 / (4\pi n_e)]^{1/3}$, $n_e$ is the electron density and $a_B$ is the Bohr radius.   
In our MFPIMC simulations we used the Metropolis algorithm  \cite{zamalin1977monte,EbelForFil,ForFilLarEbl} 
and varied both the particle number 
in the range $N_e=30 {,}  \dots {,} 50$, and the number of high-temperature factors in the range  $M=20\dots 30$. 
Bigger values of $M$ and $N_e$ are presently not  feasible.  
To analyze the influence of interparticle interaction on the exchange matrix we compare internal energy 
and pair distribution functions (PDFs) for the matrix $\|\phi_{kt}\|$ (Eq.~(\ref{eq:phikt})) with its approximation 
given by expression 

\begin{equation}
\tilde{\phi}_{kt}\approx e^{-\pi \left|x_k-x_t\right|^2}.
\label{eq:phitilda}
\end{equation}

\begin{figure}[htp]
	%\hspace{.2cm}
	%\\\includegraphics[width=8.1cm,clip=true]{lgget05.eps}
	%\includegraphics[width=8.1cm,clip=true]{gmet05.eps} 
	%[0.4cm]
%	\hspace{.2cm}
	\includegraphics[width=7.5cm,clip=true]{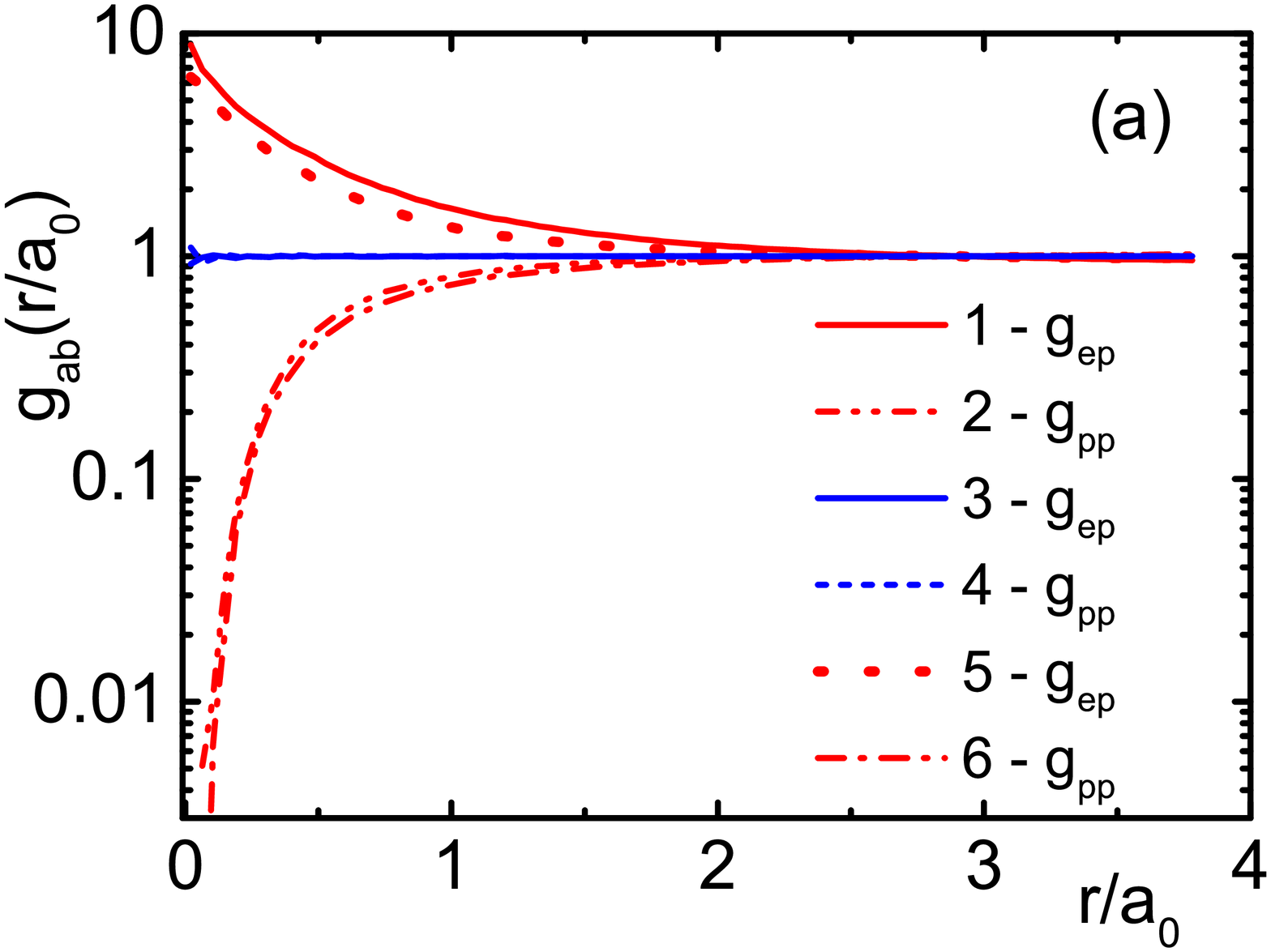}
	\includegraphics[width=7.5cm,clip=true]{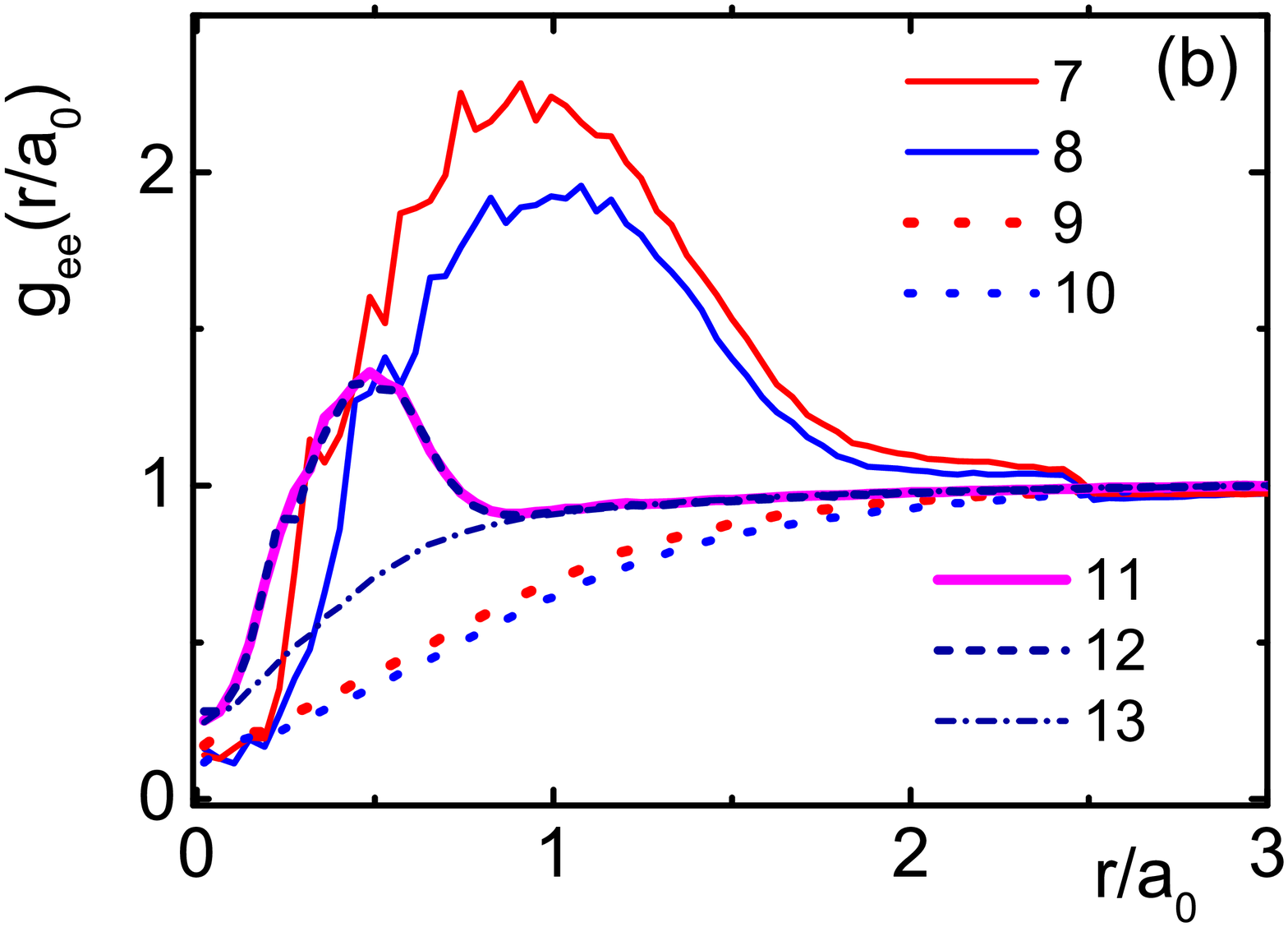}
	%\includegraphics[width=8.1cm,clip=true]{gmet125.eps}
%	\hspace{.2cm}
	\caption{(Color online)  %(Plots (a) and (b)) 
		PDFs of TCP and UEG at $r_s=2$ obtained according to Eq.~(\ref{pathint_wignerfunctionint3}) with different matrix $\psi$, 
		(a) $g_{ep}$ and $g_{pp}$, (b) $g_{ee}$. Expression (\ref{eq:phikt}) for $\psi$: red lines 1, 6, 7---for TCP  at $T/Ry=3$; red line 11---for TCP at $T/Ry=12$; 
		blue lines 3, 4, 8---for UEG at $T/Ry=3$; blue line 12---for UEG at $T/Ry=12$ (blue line 12). 
		Approximation (\ref{eq:phitilda}) for $\psi$: red lines 2, 5, 9--- for TCP at $T/Ry=3$; blue line 10---for UEG at $T/Ry=3$; blue line 13---for UEG at $T/Ry=12$. 
		Electron thermal wavelength: $\lambda_e \approx 2 a_0$ at $T/Ry=3$ ($n_e \lambda_e^3=0.256$ ) and  $\lambda_e \approx 1 a_0$ at $T/Ry=12$ ($n_e \lambda_e^3=0.032$ ).
		\label{fig:upol}
	}
\end{figure}

\textbf{Pair distribution functions.}
Let us start from the physical analysis of the spatial arrangement of electrons and positive particles,  
by studying PDFs $g_{ab}(R)$ defined as: 
%%----------------------------------------------------------------------------------------------------------------
%\begin{multline}
\begin{eqnarray}
\label{g-def} 
g_{ab}(|{\bf R}_1-{\bf R}_2|) =
\left(\frac{V}{N}\right)^2 %\left(N,V,\beta\right) \text{,}}
\sum_{\sigma}
\sum_{i,j,i\neq j}
\delta_{a_i,a}\, \delta_{a_j,b} \nonumber\\
{} \times \frac{1}{Z} \int \rm dr\; \delta({\bf R}_1-{\bf r}_i)\,\delta({\bf R}_2-{\bf r}_j)\;
\,\rho(x, \sigma ;\beta), 
\end{eqnarray}
%\end{multline}
where $a_i$ and $b_j$ are the types of particles. PDF depends only on the difference of particles coordinates $R=|{\bf R}_1-{\bf R}_2|$ because of the translational invariance of the system. The product $R^2 g_{ab}(R)$ is proportional 
%(up to a constant factor) 
to the probability to find two particles at a distance $R$. 
The PDFs averaged over spin of electrons are shown in Fig.~\ref{fig:upol}(a). 
%for a temperature $T = 12 \text{Ry}$ and density related to $r_s=2$.
In a non-interacting classical system $g_{ep}(R) = g_{pp} \equiv 1$, whereas for TCP interparticle 
interactions and  quantum statistics result in a redistribution of particles   
(see lines 1, 2, 5, 6 in Fig.~\ref{fig:upol}(a)). 
% $g_{ee}(R)\le 1$, $g_{pp}(R)\le 1$ and $g_{ep}(R)\ge 1$.  
For UEG the rigid neutralizing background is simulated as an ideal gas of uncorrelated 
classical positive charges uniformly distributed in space, so $g_{pp}$ and $g_{ep}$ are identically equal 
to unity, what is demonstrated by lines 3 and 4 in Fig.~\ref{fig:upol}(a).   

For quantum electrons $g_{ee}$ demonstrates a drastic difference in behavior (Fig.~\ref{fig:upol}(b))
due to the interference effects of the Coulomb and exchange interactions of electrons.   
At a temperature  $T/Ry=3$ for TCP and UEG lines 7 and 8 show $g_{ee}$ 
according to  Eq.~(\ref{eq:phikt})  for the matrix $\|\phi_{kt}\|$,  
while lines 9 and 10 present $g_{ee}$ corresponding to approximation $\|\tilde{\phi}_{kt}\|$ (Eq.~(\ref{eq:phitilda})). 
% for matrix $\psi$. 
For a higher temperature $T/Ry=12$ lines 11 and 12 correspond to Eq.~(\ref{eq:phikt}), while line 13 corresponds to approximation (\ref{eq:phitilda}).   

An interesting result of MFPIMC calculations is the peak in the $g_{ee}$ at interparticle distance of the order of the electron 
thermal wavelength. This peak can be associated with exchange excitons \cite{Weisskopf:PR:1939,lowdin1955quantum,lowd} 
arising due to the occurrence of the exchange--correlation positively charged  hole 
(see blue and red $g_{ep}$ in Fig.~\ref{fig:upol}(a))  and the excluded volume effect \cite{barker1972theories}. 
Contrary to approximation (\ref{eq:phitilda}) expression (\ref{eq:phikt})  more accurately accounts for 
the Coulomb and Fermi repulsion of electrons resulting in the occurrence of the peak in $g_{ee}$. 
Small oscillations of the PDF are caused by the Monte-Carlo statistical error. 

\textbf{Internal energy.}
%\subsection{Internal energy}
Calculations of the internal energy has been done by the energy  estimator 
\cite{filinov2015fermionic,filinov2020uniform}). 
At $r_s=2$ and $T/Ry = 3$ we have the following results for internal energy 
according to Eq.~(\ref{pathint_wignerfunctionint3}) with expression  A1 for $\psi$:  \\
the TCP energy $E_{tcp}/k_BT=1.270$;\\
the UEG $E_{ueg}/k_BT=1.171$. \\
For internal energy calculated according to Eq.~(\ref{pathint_wignerfunctionint3}) with approximation A2 for $\psi$ we have obtained:  \\
$E_{tcp}/k_BT=1.039$;\\
$E_{ueg}/k_BT=0.9288$.\\ 
Internal energy both for TCP and UEG is lower for the exchange matrix $\psi$ with approximation A2 than with expression A1. 
This is because the PDFs in case of approximation A1 (lines 7 and 8 in Fig.~\ref{fig:upol}(b)) are always higher than 
those in case of approximation A2 (lines 9 and 10 in Fig.~\ref{fig:upol}(b)). The difference is noticeable but not very 
big for integral thermodynamic characteristics 
as the product $R^2 g_{ab}(R)$ reduces contributions at small interparticle distances. 

\textbf{Discussion.}
Approximation (\ref{eq:phitilda}) is valid for a highly degenerate system \cite{filinov2020uniform}. In this case the determinant $\psi(x)$ is equivalent to the Gram determinant  \cite{gantmacher2005applications}  
${\rm det}\| \langle \chi(x_t) | \chi(x_k)  \rangle \|  $ of 
linearly independent system of $N_e$ vectors 
$|\chi_p(x_k)\rangle = |e^{\rm i \langle p|x_k  \rangle / \hbar }\rangle$  with the scalar product 
defined as $\langle \chi(x_t) | \chi(x_k)  \rangle =  
\int dp e^{\rm i \langle p|x_k-x_t \rangle / \hbar } e^{-\beta p^2/2m} / \int dp e^{-\beta p^2/2m} 
= e^{-\pi \left|r_{kt}\right|^2/\lambda_e^2} $.  
The  Gram determinant is always non-negative \cite{gantmacher2005applications}  
and allows to solve the \emph{``sign problem''} in PIMC simulations of fermions in the discussed approximation. 

For comparison let us consider similar matrix elements  
$e^{-{\pi} \left|x_k^{(M)} -x_t^{(0)}\right|^2} $ 
in the exchange determinants $\det\Psi(x)$ in Eq.~(\ref{rho_s}) used in standard PIMC.  
These matrix elements are independent of any potential. 
Let us note that $\psi(x)$ in Eq.~(\ref{pathint_wignerfunctionint3}) has a different structure in comparison with the matrix elements in  $\Psi(x)$, which are 
defined by the analogous scalar product but for the other set of $2N_e$  vectors 
$|e^{\rm i \langle p|x_k^{(M)}  \rangle / \hbar }\rangle$ 
and $|e^{\rm i \langle p|x_k^{(0)}  \rangle / \hbar }\rangle$.  
The exchange determinant $\det\Psi(x)$ is a sign alternating function, that transfers 
the ``sign problem'' in Eq.~(\ref{rho_s}) to a higher level. As a result the accuracy of PIMC calculations decreases 
in comparison with the approximation (\ref{eq:phitilda}). 

The MFPIMC method for the exchange matrix $\psi$ with $\phi_{kt}$ given by Eq.~(\ref{eq:phikt}) contrary to the less accurate approximation (\ref{eq:phitilda}) demonstrates 
a short--range quantum ordering of electrons arising from the Coulomb interaction of electrons with exchange holes. 
This fine phenomenon is known in the literature \cite{Weisskopf:PR:1939,lowdin1955quantum} and is associated with 
the formation of exchange excitons which modify pair correlations.
%
%%%%%\section{Acknowledgements}
\acknowledgments
We acknowledge stimulating discussions with Prof. M. Bonitz. 
%, T. Schoof, S. Groth and T. Dornheim. 
The theoretical approach to the basic equations and the algorithmic realization of the MFPIMC approach was supported 
by the Russian Science Foundation, Grant No.~20-42-04421. 
%(RNF)
%This research was supported by The Ministry of Science and Higher Education of the Russian Federation 
%(Agreement with Joint Institute for High Temperatures RAS No 075-15-2020-785 dated September 23, 2020).  
%(Fortov, судсидии)
The numerical calculations were supported by the Ministry of Science and Higher Education of the Russian Federation 
(State Assignment No. 075-00892-20-01).
%\newpage 
\section{Appendix. Derivation  of the new path integral represetation of the density matrix}
For simplicity let us prove Eq.~(2) for spinless fermions. 
The generalization for fermions with spin degree of freedom is trivial. 
The first three lines of Eq.~(2) have been obtained in \cite{larkin2017pauli,larkin2017peculiarities,ForFilLarEbl} 
after replacing variables 
$x^{(m)}$ by $q^{(m)}$ for any given permutation $P$  by the substitution 
$	x^{(m)} = (Px - x)\frac{m}{M}+x + q^{(m)}$ 
in the the standard expression for the path integral representation of density matrix like in Eq.~(\ref{rho_s}) 
\cite{feynmanquantum,zamalin1977monte}.  

Let us derive the new path integral represetation of the density matrix.    
%for the region of intermediate degeneracy which simplifies the exchange determinant calculations.  
From the expression for potential energy  $\epsilon U\bigl((P x -x)\frac{m}{M}+x + q^{(m)}	\bigr) $ 
in Eq.~(2) 
%Eq.~(\ref{pathint_wignerfunctionint5}) 
it follows that each coordinate of vector $x_k$ may be changed by  
the term $(x_{Pk} -x_k)\frac{m}{M}$ only  if $Pk \neq k$. 
The pair potentials relating to the identical and pair permutations   
will be denoted further by symbols $\Phi_{kt}=\epsilon \Phi (|r_{kt} + q_{kt}^{(m)} | )$ and 
$\tilde{\Phi}_{kt}=\epsilon \Phi (|r_{tk}\frac{2m}{M} +r_{kt} + q_{kt}^{(m)}|  )$ respectively. 
The difference of the matrix elements $\Delta \tilde{\Phi}_{kt}=\tilde{\Phi}_{kt}-\Phi_{kt}$  is non-zero only 
if the permutation changes  the number $k$ or/and $t$ ($Pk \neq k$ or/and $Pt \neq t$).  
So the non-zero elements  $\Delta \tilde{\Phi}_{kt}$ of any given permutation $P$ form 
rows and columns with correspondent  numbers $k$ and $t$. 
  
In the sum over permutations in Eq.~(2) every permutation $P$ is comprised of 
%Eq.~(\ref{pathint_wignerfunctionint5}) 
cyclic ones without common components \cite{gantmacher2005applications}, 
so without the loss of generality we can consider only cyclic permutations. In its turn the cyclic permutation 
can be presented as the composition of pair transpositions $(k,t)$  
\cite{gantmacher2005applications} like $ \phi_{12} \phi_{21} $  in Eq.~(\ref{pathint2}).

At this point we introduce an approximation allowing to reduce the ``fermionic sign problem''  and to make use of the advantages of 
direct methods of linear algebra to calculate determinants describing the exchange interaction. 
We have replaced the pair potentials $\tilde{\Phi}_{kt}$ with $\Phi_{kt}$ in row $k$ and column $t$ except their 
intersection marked by stars  $\star \tilde{\Phi}_{kt}$. 
% and $ \star \tilde{\Phi}_{tk}$. 
So in the considered approximation in row $k$ 
and column $t$ all matrix elements  $\Delta \tilde{\Phi}_{kt}=\tilde{\Phi}_{kt}-\Phi_{kt}$    are 
equal to zero except  $\star \Delta \tilde{\Phi}_{kt}$. 
% and $ \star \Delta \tilde{\Phi}_{tk}$. 
As example, tables 1 and 2 show this values for the simple  permutation comprising from two 
non-trivial cyclic permutations and four identical cycles $P_c=(1,3,5),(4,7),(2,2),(6,6),(8,8),(9,9)$ for $9$ particles. 
%As a simple example one can takes the permutation    
%comprising from two non-trivial cyclic permutations and four identical cycles $P_c=(1,3,5),(4,7),(2,2),(6,6),(8,8),(9,9)$ for $9$ particles. 
% and $\star \Delta \tilde{\Phi}_{kt}$.    
As a result of this approximation we have for $P_c$: 
%\begin{multline}
\begin{eqnarray}
&&\exp\Bigl[	-\pi \frac{|P_c x-x|^2}{M} 
-\Delta U_P{_c} \Bigr] 
\approx \phi^{*}_{13}\phi^{*}_{35}\phi^{*}_{51}\phi^{*}_{47}\phi^{*}_{74},
\nonumber
\label{pathint_wignerfunctionint6}
\end{eqnarray} 
where $ \phi^{*}_{kt}=  \exp \{-{\pi} \left|r_{kt}\right|^2/M \} \exp \{-\frac{1}{2}\sum\limits_{m = 0}^{M-1} 
\big( \epsilon\ {\star\Delta \tilde{\Phi}_{kt}} \big)\}      
$, $\phi^{*}_{kk}=1$. (The product $\phi^{*}_{31}\phi^{*}_{15}\phi^{*}_{53}$ will be accounted for by symmetric 
cycle like in $3\times3$ matrix).
% and the matrx elements related to the cycle $P_c$ are marked by stars  $\star \tilde{\Phi}_{kt}$ 
%in Tables 1, 2, 3. 
%The elements on the intersections of  
%will be market by the stars  $\star \Delta \tilde{\Phi}_{kt}$. 

The sum over all permutations gives 
the determinant  $\det\|\psi(x)\|$ of matrix $\|\phi_{kt}\|$  in Eq.~(\ref{pathint_wignerfunctionint3}). 
%Eq.~(\ref{pathint_wignerfunctionint5}).   
From the physical point of view the determinant describes the interference effects of the Coulomb and the exchange 
interactions of electrons. Table 3 (as the final result) presents the matrix of all pair interactions related 
to the cyclic permutation $P_c$.  

Due to this approximation  the particles from any given cycle interact with other particles  
with the pair potentials $\Phi_{kt}$ instead of $\tilde{\Phi}_{kt}$, while    
along the cyclic path  the particle interaction is described by $ \tilde{\Phi}_{kt}$. 
At this perturbation in potential energy ($\Phi_{kt}$ instead of $\tilde{\Phi}_{kt}$) the change in corresponding free energy is of order 
$ n_e \lambda_e^3/2  \int_V \rm d r_{12} \int  dq_{12}^{(1)}, \dots, dq_{12}^{(M-1)} 
g_{ee}(r_{12},q_{12}^{(1)} \dots q_{12}^{(M-1)}) \\
(exp(-\sum\limits_{m = 0}^{M-1} \epsilon \Delta \tilde{\Phi}_{12})-1) 
\sim n_e \lambda_e^3/2 \, g_{ee}((n_e \lambda_e^3)^{-1/3}) \\  
(-\beta \Delta \tilde{\Phi}((n_e \lambda_e^3)^{-1/3}))
\sim n_e \lambda_e^3/2 \, g_{ee}((n_e \lambda_e^3)^{-1/3}) (\beta e^2/\lambda_e)
$ 
and can be neglected for small $\Delta \tilde{\Phi}$  and $n_e \lambda_e^3 \lsim 1 $. 
Here $ \beta e^2/\lambda_e=\sqrt{\beta \rm Ry/\pi} $ and $g_{ee} $ is the pair distribution function.  
This estimation can be 
obtained within the generalization of the Mayer expansion technique for strongly coupled systems of particles 
(by the so-called algebraic approach) \cite{ruelle1999statistical,zelener1981perturbation}.  

The number and length of the cycles giving the main contribution to the thermodynamics of the fermionic system 
depends on the degeneracy and can be very large or small. At the moderate degeneracy ($n_e \lambda_e^3 \lsim 1 $)
because of the Fermi repulsion the main contribution to the exchange determinant is given by the pair permutations 
\cite{larkin2017pauli,larkin2017peculiarities}. It is a lucky case that 
the exchange matrix elements $\phi^{*}_{kt}$ and the approximate density matrix  
(\ref{pathint_wignerfunctionint3}) coincide  with the exact analogous functions for two fermions respectively
(see the  Eq.~(\ref{pathint2}) where $\mathrm{det}\bigl\|\psi(x)\bigr\| =1-\phi_{12}\phi_{21}$). 

More over under this conditions ($n_e \lambda_e^3 \lsim 1 $) for  arbitrary number of fermions 
the Laplace expansion of the exchange determinant in terms of minors and the exact pair exchange blocks 
shows that  the exchange determinant in Eq.~(\ref{pathint_wignerfunctionint3})  is approaching the product 
of the corresponding  two particle determinants $\prod (1-\phi_{kt}\phi_{tk})$  (see Eq.~(\ref{pathint2})). 
This supports the correctness of the  approximate density matrix  
(\ref{pathint_wignerfunctionint3}) and enables to carry out reliable calculations of the thermodynamic values.

%\bibliographystyle{iopart-num} 
%\bibliographystyle{eplbib} 
%\bibliography{ueg.bib}
%EXAMPLE
%\begin{thebibliography}{0}
%
%	\bibitem{b.a}
%	\Name{Author F., Author S. \and Author T.}
%	\REVIEW{Some Rev. A}{69}{1969}{9691}.
%	
%	\bibitem{b.b}
%	\Name{Author F. \and Author S.}
%	\Book{Some Book of Interest}
%	\Editor{A. Editor}
%	\Vol{9}
%	\Publ{Publishing house, City}
%	\Year{1939}
%	\Page{666}.
%	
%	\bibitem{b.c}
%	\Editor{Editor A.}
%	\Book{Some Book of Interest}
%	\Vol{9}
%	\Publ{Publishing house, City}
%	\Year{1939}
%	\Section{A}.
%	
%\end{thebibliography}

\newpage
  \begin{table}[htb]\label{tab1}
\smallskip\noindent
  	\caption{ 
  		Initial matrix of pair potential for $P_c$}
\resizebox{\linewidth}{!}{  	
  	\begin{tabular}{|c|c|c|c|c|c|c|c|c|}
  		\hline		
  		$0$ & $\tilde{\Phi}_{12}$ & $ \star \tilde{\Phi}_{13}$ & $ \tilde{\Phi}_{14}$ & $ \star \tilde{\Phi}_{15}$ & $ \tilde{\Phi}_{16}$ & $ \tilde{\Phi}_{17}$ & $ \tilde{\Phi}_{18}$& $ \tilde{\Phi}_{19}$\\
  		\hline
  		$ \tilde{\Phi}_{21}$ & $0$ & $ \tilde{\Phi}_{23}$ & $ \tilde{\Phi}_{24}$ & $ \tilde{\Phi}_{25}$ & $\Phi_{26}$ & $ \tilde{\Phi}_{27}$ & $\Phi_{28}$ & $\Phi_{29}$\\
  		\hline	
  		$ \star  \tilde{\Phi}_{31}$ & $ \tilde{\Phi}_{32}$ & $0$ & $ \tilde{\Phi}_{34}$ & $\star  \tilde{ \Phi}_{35}$ & $ \tilde{\Phi}_{36}$ & $ \tilde{\Phi}_{37}$ & $ \tilde{\Phi}_{38}$ & $ \tilde{\Phi}_{39}$\\
  		\hline
  		$ \tilde{\Phi}_{41}$ & $ \tilde{\Phi}_{42}$ & $ \tilde{\Phi}_{43}$ & $0$ & $ \tilde{\Phi}_{45}$ & $ \tilde{\Phi}_{46}$ & $\star  \tilde{\Phi}_{47}$ & $ \tilde{\Phi}_{48}$ & $ \tilde{\Phi}_{49}$\\
  		\hline
  		$\star  \tilde{\Phi}_{51}$ & $ \tilde{\Phi}_{52}$ & $ \star \tilde{\Phi}_{53}$ & $ \tilde{\Phi}_{54}$ & $0$ & $ \tilde{\Phi}_{56}$ & $ \tilde{\Phi}_{57}$ & $ \tilde{\Phi}_{58}$ & $ \tilde{\Phi}_{59}$\\
  		\hline
  		$ \tilde{\Phi}_{61}$ & $\Phi_{62}$ & $ \tilde{\Phi}_{63}$ & $ \tilde{\Phi}_{64}$ & $ \tilde{\Phi}_{65}$ & $0$ & $ \tilde{\Phi}_{67}$ & $\Phi_{68}$ & $\Phi_{69}$\\
  		\hline
  		$\tilde{\Phi}_{71}$ & $ \tilde{\Phi}_{72}$ & $ \tilde{\Phi}_{73}$ & $\star  \tilde{\Phi}_{74}$ & $ \tilde{\Phi}_{75}$ & $ \tilde{\Phi}_{76}$ & $0$ & $ \tilde{\Phi}_{78}$ & $ \tilde{\Phi}_{79}$\\
  		\hline
  		$ \tilde{\Phi}_{81}$ & $\Phi_{82}$ & $ \tilde{\Phi}_{83}$ & $ \tilde{\Phi}_{84}$ & $ \tilde{\Phi}_{85}$ & $\Phi_{86}$ & $ \tilde{\Phi}_{87}$ & $0$ & $\Phi_{89}$\\
  		\hline
  		$ \tilde{\Phi}_{91}$ & $\Phi_{92}$ & $ \tilde{\Phi}_{93}$ & $ \tilde{\Phi}_{94}$ & $ \tilde{\Phi}_{95}$ & $\Phi_{96}$ & $ \tilde{\Phi}_{97}$ & $\Phi_{89}$ & $0$\\
  		\hline	
  	\end{tabular}}
  \end{table} 
%\newpage
%\begin{center}
  \begin{table}[htb]\label{tab2}
  	\caption{ 
  		Approximation of matrix of the pair potential differences for $P_c$}
\resizebox{\linewidth}{!}{   	
  	\begin{tabular}{|c|c|c|c|c|c|c|c|c|}
  		\hline		
  		$0$ & $0$ & $ \star\Delta \tilde{\Phi}_{13}$ & $0$ & $ \star\Delta \tilde{\Phi}_{15}$ & $0$ & $0$ & $0$& $0$\\
  		\hline
  		$0$ & $0$ & $0$ & $0$ & $0$ & $0$ & $0$ & $0$ & $0$\\
  		\hline	
  		$ \star\Delta \tilde{\Phi}_{31}$ & $0$ & $0$ & $0$ & $\star \Delta \tilde{ \Phi}_{35}$ & $0$ & $0$ & $0$ & $0$\\
  		\hline
  		$0$ & $0$ & $0$ & $0$ & $0$ & $0$ & $\star \Delta \tilde{\Phi}_{47}$ & $0$ & $0$\\
  		\hline
  		$\star \Delta \tilde{\Phi}_{51}$ & $0$ & $ \star\Delta \tilde{\Phi}_{53}$ & $0$ & $0$ & $0$ & $0$ & $0$ & $0$\\
  		\hline
  		$0$ & $0$ & $0$ & $0$ & $0$ & $0$ & $0$ & $0$ & $0$\\
  		\hline
  		$0$ & $0$ & $0$ & $\star \Delta \tilde{\Phi}_{74}$ & $0$ & $0$ & $0$ & $0$ & $0$\\
  		\hline
  		$0$ & $0$ & $0$ & $0$ & $0$ & $0$ & $0$ & $0$ & $0$\\
  		\hline
  		$0$ & $0$ & $0$ & $0$ & $0$ & $0$ & $0$ & $0$ & $0$\\
  		\hline
  	\end{tabular}}
  \end{table}
%\end{center}
%\newpage
%\begin{center}
\begin{table}[htb]\label{tab3}
	\caption{ 
		Final matrix of pair potentials $\Phi_{kt}$ and $\star \tilde{\Phi}_{kt}$  for $P_c$}
\resizebox{\linewidth}{!}{ 	
	\begin{tabular}{|c|c|c|c|c|c|c|c|c|}
		\hline		
		$0$ & $\Phi_{12}$ & $ \star \tilde{\Phi}_{13}$ & $\Phi_{14}$ & $ \Phi_{15}$ & $\Phi_{16}$ & $\Phi_{17}$ & $\Phi_{18}$& $\Phi_{19}$\\
		\hline
		$\Phi_{21}$ & $0$ & $\Phi_{23}$ & $\Phi_{24}$ & $\Phi_{25}$ & $\Phi_{26}$ & $\Phi_{27}$ & $\Phi_{28}$ & $\Phi_{29}$\\
		\hline	
		$ \Phi_{31}$ & $\Phi_{32}$ & $0$ & $\Phi_{34}$ & $\star  \tilde{ \Phi}_{35}$ & $\Phi_{36}$ & $\Phi_{37}$ & $\Phi_{38}$ & $\Phi_{39}$\\
		\hline
		$\Phi_{41}$ & $\Phi_{42}$ & $\Phi_{43}$ & $0$ & $\Phi_{45}$ & $\Phi_{46}$ & $\star  \tilde{\Phi}_{47}$ & $\Phi_{48}$ & $\Phi_{49}$\\
		\hline
		$\star  \tilde{\Phi}_{51}$ & $\Phi_{52}$ & $ \Phi_{53}$ & $\Phi_{54}$ & $0$ & $\Phi_{56}$ & $\Phi_{57}$ & $\Phi_{58}$ & $\Phi_{59}$\\
		\hline
		$\Phi_{61}$ & $\Phi_{62}$ & $\Phi_{63}$ & $\Phi_{64}$ & $\Phi_{65}$ & $0$ & $\Phi_{67}$ & $\Phi_{68}$ & $\Phi_{69}$\\
		\hline
		$\Phi_{71}$ & $\Phi_{72}$ & $\Phi_{73}$ & $\star  \tilde{\Phi}_{74}$ & $\Phi_{75}$ & $\Phi_{76}$ & $0$ & $\Phi_{78}$ & $\Phi_{79}$\\
		\hline
		$\Phi_{81}$ & $\Phi_{82}$ & $\Phi_{83}$ & $\Phi_{84}$ & $\Phi_{85}$ & $\Phi_{86}$ & $\Phi_{87}$ & $0$ & $\Phi_{89}$\\
		\hline
		$\Phi_{91}$ & $\Phi_{92}$ & $\Phi_{93}$ & $\Phi_{94}$ & $\Phi_{95}$ & $\Phi_{96}$ & $\Phi_{97}$ & $\Phi_{98}$ & $0$\\
		\hline
	\end{tabular}}
\end{table}
%\end{center}

%\newpage
%\bibliographystyle{eplbib} 
%\bibliography{ueg.bib}

\begin{thebibliography}{10}
	\expandafter\ifx\csname url\endcsname\relax\def\url#1{\texttt{#1}}\fi
	
	\bibitem{feynmanquantum}
	\Name{Feynman R.~P., Hibbs A.~R. \and Styer D.~F.} \Book{Quantum mechanics and
		path integrals} (Courier Corporation) 2010.
	
	\bibitem{zamalin1977monte}
	\Name{Zamalin V., Norman G. \and Filinov V.} \Book{The monte carlo method in
		statistical thermodynamics} (1977).
	
	\bibitem{EbelForFil}
	\Name{Ebeling W., Fortov V. \and Filinov V.} \Book{Quantum Statistics of Dense
		Gases and Nonideal Plasmas} (Springer, Berlin) 2017.
	
	\bibitem{ForFilLarEbl}
	\Name{Fortov V., Filinov V., Larkin A. \and Ebeling W.} \Book{Statistical
		physics of Dense Gases and Nonideal Plasmas} (PhysMatLit, Moscow) 2020.
	
	\bibitem{larkin2017pauli}
	\Name{Larkin A., Filinov V. \and Fortov V.} \REVIEW{Contributions to Plasma
		Physics}{57}{2017}{506}.
	
	\bibitem{larkin2017peculiarities}
	\Name{Larkin A., Filinov V. \and Fortov V.} \REVIEW{Journal of Physics A:
		Mathematical and Theoretical}{51}{2017}{035002}.
	
	\bibitem{brown2013path}
	\Name{Brown E.~W., Clark B.~K., DuBois J.~L. \and Ceperley D.~M.}
	\REVIEW{Physical review letters}{110}{2013}{146405}.
	
	\bibitem{dornheim2018uniform}
	\Name{Dornheim T., Groth S. \and Bonitz M.} \REVIEW{Physics
		Reports}{744}{2018}{1}.
	
	\bibitem{Weisskopf:PR:1939}
	\Name{Weisskopf V.~F.} \REVIEW{Physical Review}{56}{1939}{72}.
	
	\bibitem{lowdin1955quantum}
	\Name{L{\"o}wdin P.-O.} \REVIEW{Physical review}{97}{1955}{1509}.
	
	\bibitem{lowd}
	\Name{Himpsel F.} \REVIEW{arXiv preprint arXiv:1701.08080}{}{2017}{}.
	
	\bibitem{filinov2020uniform}
	\Name{Filinov V., Larkin A. \and Levashov P.} \REVIEW{Physical Review
		E}{102}{2020}{033203}.
	
	\bibitem{filinov2015fermionic}
	\Name{Filinov V., Fortov V., Bonitz M. \and Moldabekov Z.} \REVIEW{Physical
		Review E}{91}{2015}{033108}.
	
	\bibitem{barker1972theories}
	\Name{Barker J. \and Henderson D.} \REVIEW{Annual review of physical
		chemistry}{23}{1972}{439}.
	
	\bibitem{gantmacher2005applications}
	\Name{Gantmacher F.~R. \and Brenner J.~L.} \Book{Applications of the Theory of
		Matrices} (Courier Corporation) 2005.
	
	\bibitem{ruelle1999statistical}
	\Name{Ruelle D.} \Book{Statistical mechanics: Rigorous results} (World
	Scientific) 1999.
	
	\bibitem{zelener1981perturbation}
	\Name{Zelener B., Norman G. \and Filinov V.} \Book{Perturbation theory and
		pseudopotential in statistical thermodynamics} (Nauka,Moscow) 1981.
	
\end{thebibliography}
%
%% here a revision
%%
%%\revision{Insert here the text.
%%See fig.~\ref{fig.1}, table~\ref{tab.1} and eq.~(\ref{eq.1}).
%%See also~\cite{b.a,b.b}.}
%%
%%here a shortcut $\emc$ and again $\emc$
%%
%%
%%\begin{equation}
%%\label{eq.1}
%%0\neq1
%%\end{equation}
%%
%%\begin{figure}
%%\onefigure{epl-template.eps}
%%\caption{Figure caption.}
%%\label{fig.1}
%%\end{figure}
%%
%%
%%\begin{table}
%%\caption{Table caption.}
%%\label{tab.1}
%%\begin{center}
%%\begin{tabular}{lcr}
%%first  & table & row\\
%%second & table & row
%%\end{tabular}
%%\end{center}
%%\end{table}

%%\acknowledgments
%%Insert here the text.

\end{document}